\documentclass[12pt,a4paper,oneside,final,tilepage,onecolumn,thmsal]{article}
\usepackage[dvips]{graphics}
\usepackage{epsfig}
\usepackage{amsmath}

\newcommand{\QED}{\hspace*{\fill}\rule{2.5mm}{2.5mm}}

\usepackage{graphics}
\begin{document}
\def\beq{\begin{equation}}
\def\eeq{\end{equation}}
\def\bea{\begin{eqnarray}}
\def\eea{\end{eqnarray}}
\def\ve{\vert}
\def\vel{\left|}
\def\ver{\right|}
\def\nnb{\nonumber}
\def\ga{\left(}
\def\dr{\right)}
\def\aga{\left\{}
\def\adr{\right\}}
\def\rar{\rightarrow}
\def\nnb{\nonumber}
\def\la{\langle}
\def\ra{\rangle}
\def\ba{\begin{array}}
\def\ea{\end{array}}
\def\tep{$B \rar K \ell^+ \ell^-$}
\def\tepm{$B \rar K \mu^+ \mu^-$}
\def\tept{$B \rar K \tau^+ \tau^-$}
\def\ds{\displaystyle}
\title{{\small {\bf Fourth generation effects on the rare
$B\rightarrow K^{*}\nu\bar{\nu}$ decay }}}
\author{\vspace{1cm}\\
{\small T. BARAKAT} \thanks {electronic address:
barakat@ciu.edu.tr}\\ {\small Engineering Faculty, Cyprus
International University},\\ {\small Lefko\c{s}a, Mersin 10 -
Turkey } }
\date{}
\begin{titlepage}
\maketitle
\thispagestyle{empty}
\begin{abstract}
\baselineskip .8 cm If the fourth generation fermions exist, the
new quarks could influence the branching ratio of the rare
$B\rightarrow K^{*}\nu\bar{\nu}$ decay. Two possible solutions of
the fourth generation CKM factor
$(V^{*}_{\acute{t}s}V_{\acute{t}b})^(\pm)$ are obtained as a
function of the new $\acute{t}$-quark mass $(m_{\acute{t}})$ from
the experimental results of the $B\rightarrow X_{s}\gamma$
together with the semileptonic decay $B\rightarrow
X_{c}\ell\bar{\nu}$. The branching ratio of the decay
$B\rightarrow K^{*}\nu\bar{\nu}$ in the two cases is estimated. In
one case, a significant enhancement to the branching ratio of the
decay $B\rightarrow K^{*}\nu\bar{\nu}$ over the SM is recorded,
while results are almost same in another case. If a fourth
generation should exist in nature and nature chooses the former
case, this B meson decay could be a good probe to the existence of
the fourth generation, or perhaps a signal for a new physics.
\end{abstract}
\vspace{1cm}
%~~PACS number(s): 13.25.Gv, 13.20.--v, 13.10.+q
\end{titlepage}
\section{ Introduction}
\baselineskip .8cm \hspace{0.6cm} The SM model has been widely
discussed in the literature, and serves as an explicit model for
studying all low energy experimental data. But there is no doubt
that the SM is an incomplete theory. Among the unsolved problems
within the SM is the CP violation, and the number of generations.
In SM there are three generations, and yet, there is no
theoretical argument to explain why there are three and only three
generations in SM, and there is neither an experimental evidence
for a fourth generation nor does any experiment exclude such extra
generations. Therefore, if we believe that the fourth generation
fermions really exist in Nature, we should give their mass
spectrum, and take into account their physical effects in low
energy physic. One of the promising areas in the experimental
search for the fourth generation, via its indirect loop effects,
is the rare B meson decays.
 On this basis, serious attempts to study the effects of the fourth generation
 on the rare B meson were made by many authors. For examples, the effects of the fourth
 generation on the branching ratio of the $B \rightarrow X_{s}\ell^{+}\ell^{-}$,
and the $B \rightarrow X_{s}\gamma$ decays is analyzed in [1]. In
[2] the fourth generation effects on the rare exclusive $B
\rightarrow K^{*}\ell^{+}\ell^{-}$ decay are studied. In [3] the
contributions of the fourth generation to the $B_{s}\rightarrow
\nu \bar{\nu}\gamma$ decay is analyzed. Moreover, the introduction
of the fourth generation fermions can also affect CP violating
parameters $\acute{\epsilon}/\epsilon$ in the Kaon system [4].

It is hoped that a definite answer on possible fourth generation
at the upcoming $KEK$ and SLAC B-factories will be found, where
this year the upgraded B-factories at SLAC, and KEK will provide
us with the first experimental data. Amongst the rare flavor
changing decays, the exclusive decay $B \rightarrow
K^{*}\nu\bar{\nu}$ provokes special interest. In particular, the
SM has been exploited to establish a bound on the branching ratio
of the above-mentioned decay of the order $\sim 10^{-5}$, which
can be quite measurable for the upcoming $KEK$ and SLAC
B-factories, and they are sensitive to the various extensions to
the SM because these decays occur only through loops in the SM.
Therefore, in this work we will investigate the decay $B
\rightarrow K^{*}\nu\bar{\nu}$ in the existence of a new up-like
quark $\acute{t}$ in a sequential fourth generation model SM,
which we shall call (SM4) hereafter for the sake of simplicity.
This model is considered as natural extension of the SM, where the
fourth generation model is introduced in the same way the three
generations are introduced in the SM, so no new operators appear,
and clearly the full operator set is exactly the same as in SM.
Hence, the fourth generation will change only the values of the
Wilson coefficients via virtual exchange of the fourth generations
up-like quark $\acute{t}$.
 Subsequently, this paper is organized
as follows: in Section 2, the relevant effective Hamiltonian for
the decay $B\rightarrow K^{*}\nu\bar{\nu}$ in the existence of a
new up-like quark $\acute{t}$ in a sequential fourth generation
model (SM4) is presented; and in section 3, the dependence of the
branching ratio on the fourth generation model parameters for the
decay of interest is studied using the results of the Light- Cone
QCD sum rules for estimating form factors; and finally a brief
discussion of the results is given.
\section{Effective Hamiltonian}
\hspace{0.6cm} The matrix element of the $B\rightarrow
K^{*}\nu\bar{\nu}$ decay at quark level is described by
$b\rightarrow s\nu\bar{\nu}$ transition for whom the effective
Hamiltonian at $O(\mu)$ scale can be written as:
\begin{eqnarray}
H_{eff}&=&\frac{4G_{F}}{\sqrt{2}}V^{*}_{ts}V_{tb}\sum_{i=1}^{10}C_{i}
(\mu)O_{i}(\mu),
\end{eqnarray}
where the full set of the operators $O_{i}(\mu)$, and the
corresponding expressions for the Wilson coefficients $C_{i}(\mu)$
in the SM are given in [5]. As has been mentioned in the
introduction, no new operators appear, and clearly the full
operator set is exactly same as in SM, thus the fourth generation
changes only the values of the Wilson coefficients $C_{7}(\mu)$,
$C_{9}(\mu)$, and $C_{10}(\mu)$ via virtual exchange of the fourth
generation up quark $\acute{t}$. Therefore, the above mentioned
Wilson coefficients can be written in the following form:
\begin{eqnarray}
C^{SM4}_{7}(\mu)=C^{SM}_{7}(\mu)+\frac{V^{*}_{\acute{t}s}V_{\acute{t}b}}
{V^{*}_{ts}V_{tb}}C^{new}_{7}(\mu),
\end{eqnarray}
\begin{eqnarray}
C^{SM4}_{9}(\mu)=C^{SM}_{9}(\mu)+\frac{V^{*}_{\acute{t}s}V_{\acute{t}b}}
{V^{*}_{ts}V_{tb}}C^{new}_{9}(\mu),
\end{eqnarray}
\begin{eqnarray}
C^{SM4}_{10}(\mu)=C^{SM}_{10}(\mu)+\frac{V^{*}_{\acute{t}s}V_{\acute{t}b}}
{V^{*}_{ts}V_{tb}}C^{new}_{10}(\mu),
\end{eqnarray}
where the last terms in these expressions describe the
contributions of the $\acute{t}$ quark to the Wilson coefficients.
$V_{\acute{t}s}$, and $V_{\acute{t}b}$ are the two elements of the
$4\times 4$ Cabibbo-Kobayashi-Maskawa (CKM) matrix. In deriving
Eqs.(2-4) we factored out the term $V^{*}_{ts}V_{tb}$ in the
effective Hamiltonian given in Eq.(1). The explicit forms of the
$C^{new}_{i}$ can easily be obtained from the corresponding Wilson
coefficients in SM by simply substituting $m_{t}\rightarrow
m_{\acute{t}}$ [5,6]. Neglecting the s quark mass, the above
effective Hamiltonian leads to the following matrix element for
the $b\rightarrow s\nu\bar{\nu}$ decay in the SM [7]:
\begin{eqnarray}
H_{eff}&=&\frac{G_{F}
\alpha}{2\pi\sqrt{2}sin^{2}\theta_{w}}C^{(SM)}_{11}
V^{*}_{ts}V_{tb}\bar{s}\gamma_{\mu}(1-\gamma_{5})b\bar{\nu}\gamma_{\mu}
(1-\gamma_{5})\nu,
\end{eqnarray}
where $G_{F}$ is the Fermi coupling constant, $\alpha$ is the fine
structure constant and $V^{*}_{ts}V_{tb}$ are products of
Cabibbo-Kabayashi-Maskawa matrix elements. The resulting
expression of Wilson coefficient $C^{(SM)}_{11}$, which was
derived in the context of the SM including $O(\alpha_{s})$
corrections is [8,9]
\begin{eqnarray}
C_{11}^{(SM)}=\left[X_{0}(x)+\frac{\alpha_{s}}{4\pi}X_{1}(x)\right],
\end{eqnarray}
with
\begin{eqnarray}
X_{0}(x)= \frac{x}{8}\left[\frac{x +2}{x-1}+\frac{3(x-2)}
{(x-1)^{2}}lnx\right],
\end{eqnarray}
where $x=\frac{m^{2}_{t}}{m^{2}_{W}}$, and
\begin{eqnarray}
X_{1}(x)&=&\frac{4x^{3}-5x^{2}-23x}{3(x-1)^{2}}-
\frac{x^{4}+x^{3}-11x^{2}+x}{(x-1)^{3}}lnx+
\frac{x^{4}-x^{3}-4x^{2}-8x}{2(x-1)^{3}}ln^{2}x \nonumber \\
&+&\frac{x^{3}-4x}{(x-1)^{2}}Li_{2}(1-x)+8x\frac{\partial X_{0}(x)}{\partial x}
lnx_{\mu}.
\end{eqnarray}
Here $Li_{2}(1-x)=\int_{1}^{x}\frac{lnt}{1-t}dt$ and
$x_{\mu}=\frac{\mu^{2}}{m_{w}^{2}}$ with $\mu=O(m_{t})$.

At $\mu=m_{t}$, the QCD correction for $X_{1}(x)$ term is very
small (around $\sim 3\%$). From the theoretical point of view, the
transition $b\rightarrow s\nu\bar{\nu}$ is a very clean process,
since it is practically free from the scale dependence, and free
from any long distance effects. In addition, the presence of a
single operator governing the inclusive $b \rightarrow s
\nu\bar{\nu}$ transition is an appealing property. Therefore, the
theoretical uncertainty within the SM is only related to the value
of the Wilson coefficient $C^{(SM)}_{11}$ due to the uncertainty
in the top quark mass. In this work, we have considered possible
new physics in $b \rightarrow s \nu\bar{\nu}$ only through the
value of that of Wilson coefficient.

 In this spirit, the transition $b\rightarrow s\nu\bar{\nu}$ in Eq.(5)
 can only include extra contribution due to the fourth generation
 fermion, hence, the fourth generation fermion contribution modify only
the value of the Wilson coefficient $C^{(SM)}_{11}$ (see
Eqs.(2-4)), and it does not induce any new operators:
\begin{eqnarray}
C_{11}^{SM4}(\mu)&=&C^{(SM)}_{11}(\mu)+\frac{V^{*}_{\acute{t}s}V_{\acute{t}b}}
{V^{*}_{tb}V_{ts}}C^{(new)}(\mu),
\end{eqnarray}
where $C^{(new)}(\mu)$ can be obtained from $C^{SM}_{11}(\mu)$ by
substituting $m_{t}\rightarrow m_{\acute{t}}$.

As a result, we obtain a modified effective Hamiltonian, which
represents $b \rightarrow s \nu\bar{\nu}$ decay in the presence of
the fourth generation fermion:
\begin{eqnarray}
H_{eff}=\frac{G\alpha}{2\pi\sqrt{2}sin^{2}\theta_{w}}V^{*}_{ts}V_{tb}
[C_{11}^{(SM4)}]
\bar{s}\gamma_{\mu}(1-\gamma_{5})b\bar{\nu}\gamma_{\mu}
(1-\gamma_{5})\nu.
\end{eqnarray}
However, in spite of such theoretical advantages, it would be a
very difficult task to detect the inclusive $b \rightarrow s
\nu\bar{\nu}$  decay experimentally, because the final state
contains two missing neutrinos and many hadrons. Therefore, only
the exclusive channels, namely $B \rightarrow K^{*}(\rho)
\nu\bar{\nu}$, are well suited to search for and constrain for
possible "new physics" effects.

In order to compute $B \rightarrow K^{*} \nu\bar{\nu}$ decay, we
need the matrix elements of the effective Hamiltonian Eq.(10)
between the final and initial meson states. This problem is
related to the non-perturbative sector of QCD and can be solved
only by using non-perturbative methods. The matrix element $<K^{*}
\mid H_{eff}\mid B>$ has been investigated in a framework of
different approaches, such as chiral perturbation theory [10],
three point QCD sum rules [11], relativistic quark model by the
light front formalism [12], effective heavy quark theory [13], and
light cone QCD sum rules [14,15]. As a result, the hadronic matrix
element for the $B \rightarrow K^{*} \nu\bar{\nu}$ can be
parameterized in terms of five form factors:
\begin{eqnarray}
<K^{*}(p_{2},\epsilon) \mid \bar{s}\gamma_{\mu}(1-\gamma_{5})b\mid
B(p_{1})> = -\frac{2V(q^{2})}{m_{B}+m_{K^{*}}}
\epsilon_{\mu\nu\rho\sigma}p_{2}^{\rho}q^{\sigma}\epsilon^{*\nu}\nonumber
\\
 -i \left[\epsilon_{\mu}^{*}(m_{B}+m_{K^{*}})A_{1}(q^{2})
-(\epsilon^{*}q)(p_{1}+p_{2})_{\mu}\frac{A_{2}(q^{2})}{m_{B}+m_{K^{*}}}
\right. \nonumber \\
 - \left. q_{\mu}(\epsilon^{*}q)\frac{2m_{K^{*}}}{q^{2}}
(A_{3}(q^{2})-A_{0}(q^{2})) \right],
\end{eqnarray}
where  $\epsilon_{\mu}$, is the polarization 4-vector of $K^{*}$ meson. The
form factor $A_{3}(q^{2})$ can be written as a linear combination of the form
factors $A_{1}$ and $A_{2}$:
\begin{eqnarray}
A_{3}(q^{2})=\frac{1}{2m_{K^{*}}}\left[(m_{B}+m_{K^{*}})A_{1}(q^{2})-
(m_{B}-m_{K^{*}})A_{2}(q^{2})\right],
\end{eqnarray}
where $q=p_{1}-p_{2}$, and $A_{3}(q^{2}=0)=A_{0}(q^{2}=0)$.

After performing summation over  $K^{*}$ meson polarization and
taking into account the number of light neutrinos $N_{\nu}=3$ for
the differential decay width, we get in [7]:
\begin{eqnarray}
\frac{d\Gamma(B \rightarrow K^{*} \nu\bar{\nu})}{ds}=
\frac{G_{F}^{2}\alpha^{2}\mid
V_{tb}V^{*}_{ts}\mid^{2}}{2^{10}\pi^{5}sin^{4}\theta_{w}}
\lambda^{1/2}(1,r,s)m_{B}^{5}\mid C^{SM4}_{11} \mid^{2}\otimes
\nonumber\\
 \left\{8\lambda s\frac{V^{2}}{(1+\sqrt{r})^{2}}+
\frac{1}{r}\biggl[\lambda^{2}\frac{A_{2}}{(1+\sqrt{r})^{2}}\right.
\nonumber\\ +\left.(1+\sqrt{r})^{2}(\lambda+12rs)
A_{1}^{2}-2\lambda(1-r-s)Re(A_{1}A_{2}) \biggr] \right\},
\end{eqnarray}
where  $\lambda (1,r,s)=1+r^{2}+s^{2}-2rs-2r-2s$ is the usual
triangle function with  $r=\frac{m^{2}_{K^{*}}}{m^{2}_{B}}$  and
$s=\frac{q^{2}}{m^{2}_{B}}$. From Eq.(13), we can see that the
decay width for $B \rightarrow K^{*} \nu\bar{\nu}$ contains three
form factors: V, $A_{1}$, and $A_{2}$. These form factors were
calculated in the framework of QCD sum rules in [14,15,16].
However, in this work, in estimating the total decay width, we
have used the results of [16], where these form factors were
calculated by including one-loop radiative corrections to the
leading twist 2 contribution:
\begin{eqnarray}
F(q^{2})=\frac{F(0)}{1-a_{F}(q^{2}/m^{2}_{B})+b_{F}(q^{2}/m^{2}_{B})^{2}},
\end{eqnarray}
and the relevant values of the form factors at $q^{2}=0$ are:
\begin{eqnarray}
 A_{1}^{B \rightarrow K^{*}}(q^{2}=0)=0.35\pm 0.05,{~~} with{~~} a_{F}=0.54,
{~~}and{~~} b_{F}=-0.02,
\end{eqnarray}
\begin{eqnarray}
A_{2}^{B \rightarrow K^{*}}(q^{2}=0)=0.30\pm 0.05,{~~} with{~~} a_{F}=1.02,
{~~} and{~~} b_{F}=0.08,
\end{eqnarray}
and
\begin{eqnarray}
V^{B \rightarrow K^{*}}(q^{2}=0)=0.47\pm 0.08,{~~} with{~~} a_{F}=1.50,{~~}
and{~~} b_{F}=0.51.
\end{eqnarray}
 Note that all errors, which come out, are due to the uncertainties of the
b-quark mass, the Borel parameter variation, wave functions, and
radiative corrections are quadrature added in. Finally, to obtain
quantitative results we need the value of the fourth generation
CKM matrix elements $ V^{*}_{\acute{t}s}V_{\acute{t}b}$. For this
aim following [17], we will use the experimental results of the
decay $BR(B \rightarrow X_{s}\gamma)$ together with $BR(B
\rightarrow X_{c}e\bar{\nu_{e}})$ to determine the fourth
generation CKM factor $V^{*}_{\acute{t}s}V_{\acute{t}b}$. However,
in order to reduce the uncertainties arising from b-quark mass, we
consider the following ratio:
\begin{eqnarray}
R_{quark}=\frac{BR(B \rightarrow X_{s}\gamma)}{BR(B \rightarrow
X_{c}e\bar{\nu_{e}})}.
\end{eqnarray}
In the leading logarithmic approximation this ratio can be
summarized in a compact form as follows [18]:
\begin{eqnarray}
R_{quark}=\frac{\mid V^{*}_{ts}V_{tb} \mid ^{2}}{\mid V_{cb} \mid
^{2}}\frac{6\alpha}{\pi f(z)}  \mid C^{SM4}_{7}(m_{b}) \mid ^{2},
\end{eqnarray}
where
\begin{eqnarray}
f(z)=1-8z+8z^{3}-z^{4}-12z^{2}lnz {~~~~~~~} with {~~~}
z=\frac{m^{2}_{c,pole}}{m^{2}_{b,pole}}
\end{eqnarray}
is the phase space factor in $BR(B \rightarrow
X_{c}e\bar{\nu_{e}})$, and $\alpha= e^{2}/4\pi$. In the case of
four generation there is an additional contribution to $B
\rightarrow X_{s}\gamma$ from the virtual exchange of the fourth
generation up quark $\acute{t}$. The Wilson coefficients of the
dipole operators are given by:
\begin{eqnarray}
C^{SM4}_{7,8}(m_{b})=C^{SM}_{7,8}(m_{b})+\frac{
V^{*}_{\acute{t}s}V_{\acute{t}b}}
{V^{*}_{ts}V_{tb}}C^{new}_{7,8}(m_{b}),
\end{eqnarray}
where $C^{new}_{7,8}(m_{b})$ present the contributions of
$\acute{t}$ to the Wilson coefficients, and
$V^{*}_{\acute{t}s}V_{\acute{t}b}$ are the fourth generation CKM
matrix factor which we need now. With these Wilson coefficients
and the experiment results of the decays $BR(B \rightarrow
X_{s}\gamma)=2.66 \times 10^{-4}$, together with the semileptonic
$BR(B \rightarrow X_{c}e\bar{\nu_{e}})$=$0.103\pm 0.01$ [19,20]
decay, we obtain the results of the fourth generation CKM factor $
V^{*}_{\acute{t}s}V_{\acute{t}b}$. There exist two cases, a
positive, and a negative one [17]:
\begin{eqnarray}
( V^{*}_{\acute{t}s}V_{\acute{t}b}) ^{\pm}=\biggl[\pm \sqrt{
\frac{R_{quark} \mid V_{cb}\mid ^{2}\pi f(z)}{6\alpha \mid
V^{*}_{ts}V_{tb}\mid ^{2}}}-C^{(SM)}_{7}(m_{b}) \biggr] \frac{
V^{*}_{ts}V_{tb}}{C^{(new)}_{7}(m_{b})}.
\end{eqnarray}
The values for $V^{*}_{\acute{t}s}V_{\acute{t}b}$ are listed in
Table 1.
\begin{table}[h]
\vskip .5cm
\begin{center}
\begin{tabular}{|c|c|c|c|c|c|c|c|c} \hline
$m_{\acute{t}} (GeV)$ &50 & 100 & 150&200 &250&300&350 \\ \hline
$(V^{*}_{\acute{t}s}V_{\acute{t}b})^{+}/10^{-2}$&-11.591 &-9.259
&-8.126 &-7.501&-7.116 &-6.861&-6.580 \\
$(V^{*}_{\acute{t}s}V_{\acute{t}b})^{-}/10^{-3}$ & 3.564 &2.850
&2.502 &2.309 &2.191 & 2.113&2.205 \\ \hline \hline $m_{\acute{t}}
(GeV)$ & 400&500&600&700&800&900&1000 \\ \hline
$(V^{*}_{\acute{t}s}V_{\acute{t}b})^{+}/10^{-2}$&-6.548
&-6.369&-6.255 &-6.178&-6.123 &-6.082&-6.051 \\
$(V^{*}_{\acute{t}s}V_{\acute{t}b})^{-}/10^{-3}$ &2.016
&1.961&1.926 &1.902&1.885&1.872 &1.863\\ \hline
\end{tabular}
\caption{The numerical values of
$V^{*}_{\acute{t}s}V_{\acute{t}b}$ for different values of
$\acute{t}$ for $BR(B \rightarrow X_{s}\gamma)=2.66\times
10^{-4}$. }
\end{center}
\end{table}

A few comments about the numerical values of
$(V^{*}_{\acute{t}s}V_{\acute{t}b})^{\pm}$ are in order. From
unitarity condition of the CKM matrix we have
\begin{eqnarray}
V^{*}_{us}V_{ub}+V^{*}_{cs}V_{cb}+V^{*}_{ts}V_{tb}+V^{*}_{\acute{t}s}V_{\acute{t}b}=0.
\end{eqnarray}
If the average values of the CKM matrix elements in the SM are
used [20], the sum of the first three terms in Eq.(23) is about
$7.6\times 10^{-2}$. Substituting the value of
$(V^{*}_{\acute{t}s}V_{\acute{t}b})^{(+)}$ from Table 1, we
observe that the sum of the four terms on the left-hand side of
Eq.(22) is closer to zero compared to the SM case, since
$(V^{*}_{\acute{t}s}V_{\acute{t}b})^{(+)}$  is very close to the
sum of the first three terms, but with opposite sign. On the other
hand if we consider $(V^{*}_{\acute{t}s}V_{\acute{t}b})^{-}$,
whose value is about $ 10^{-3}$, which is one order of magnitude
smaller compared to the previous case. However, it should be noted
that the data for the CKM is not determined to very high accuracy,
and the error in sum of the first three terms in Eq.(20) is about
$\pm 0.6\times 10^{-2}$. It is easy to see then that, the value of
$(V^{*}_{\acute{t}s}V_{\acute{t}b})^{-}$ is within this error
range. In summary both $(V^{*}_{\acute{t}s}V_{\acute{t}b})^{+}$,
and $(V^{*}_{\acute{t}s}V_{\acute{t}b})^{-}$ satisfy the unitarity
condition of CKM, moreover, $\mid
(V^{*}_{\acute{t}s}V_{\acute{t}b})\mid ^{-} \leq
 10^{-1}\times \mid (V^{*}_{\acute{t}s}V_{\acute{t}b})\mid ^{+}$.
 Therefore, from our numerical analysis one cannot escape the conclusion
 that, the $(V^{*}_{\acute{t}s}V_{\acute{t}b})^{-}$ contribution to the
physical quantities should be practically indistinguishable from
SM results, and our numerical analysis confirms this expectation.
We now go on to put the above points in perspective.
\section{Numerical Analysis}
To calculate the branching ratio in SM4, and to study the
influence of the fourth generation on the branching ratio $BR(B
\rightarrow K^{*} \nu\bar{\nu})$, the following values have been
used as input parameters:\\ $G_{F}=1.17{~}.10^{-5}~ GeV^{-2}$,
$\alpha =1/137$, $m_{b}= 5.0$ GeV, $m_{B}= 5.28$ GeV, $\mid
V^{*}_{ts}V_{tb}\mid$=0.045, and the lifetime is taken as
$\tau(B_{d})=1.56\times 10^{-12}$ s [20]. For illustrative
purposes, the branching ratio (BR) for $B \rightarrow K^{*}
\nu\bar{\nu}$ decay as a function of $m_{\acute{t}}$ for its
different values of $(V^{*}_{\acute{t}s}V_{\acute{t}b})^{\pm}$ is
shown in figure 1. It can be seen that when
$V^{*}_{\acute{t}s}V_{\acute{t}b}$ take positive values, i.e.
$(V^{*}_{\acute{t}s}V_{\acute{t}b})^{-}$, the branching ratio (BR)
is almost overlap with that of SM. That is, the results in SM4 are
the same as that in SM, except a peak in the curve when
$m_{\acute{t}}$ takes values $m_{\acute{t}}\geq 210 GeV$. The
reason is not because there is new prediction deviation from SM,
but only because of the second term of Eq.(21). In this case, it
does not show the new effects of $m_{\acute{t}}$. Also, we can not
obtain the information of the existence of the fourth generation
from the branching ratio (BR) for $B \rightarrow K^{*}
\nu\bar{\nu}$, although we can not exclude them either. This is
because, from Table 1, the values
$(V^{*}_{\acute{t}s}V_{\acute{t}b})^{-}$ are positive. But they
are of order $10^{-3}$, and is very small. The values of
$V^{*}_{ts}V_{tb}$ are about ten times larger than them
$V^{*}_{ts}=0.038$, $V_{tb}=0.9995$ see ref. [20].

But in the second case, when the values of
$V^{*}_{\acute{t}s}V_{\acute{t}b}$ are negative, i.e.
$(V^{*}_{\acute{t}s}V_{\acute{t}b})^{+}$, the curve of branching
ratio (BR) for $B \rightarrow K^{*} \nu\bar{\nu}$, is quit
different from that of the SM. This can be clearly seen from
figure 1. The enhancement of the branching ratio increases rapidly
with the increasing of $m_{\acute{t}}$. In this case, the fourth
generation effects are shown clearly. The reason is that
$(V^{*}_{\acute{t}s}V_{\acute{t}b})^{+}$ is 2-3 times larger than
$V^{*}_{ts}V_{tb}$ so that the last term in Eq.(21) becomes
important, and it depends on the $\acute{t}$ mass strongly. Thus
the effect of the fourth generation is significant. In figure 2.
we show the dependence of the differential branching ratio $dBR(B
\rightarrow K^{*} \nu\bar{\nu})$/ds as functions of s; $ 0\leq s
\leq (1+\sqrt{r})^{2}$, for $m_{\acute{t}}$= 300 GeV. It can be
seen their that, when $V^{*}_{\acute{t}s}V_{\acute{t}b}$ takes
positive values, i.e. $(V^{*}_{\acute{t}s}V_{\acute{t}b})^{-}$,
the differential decay width is almost overlap with that of SM.
That is, the results in SM4 are the same as that in SM, except a
peak in the curve when $ 0.4\leq s \leq 0.6$. But in the second
case, when the values of $V^{*}_{\acute{t}s}V_{\acute{t}b}$ are
negative, i.e $(V^{*}_{\acute{t}s}V_{\acute{t}b})^{+}$. The curve
of the differential decay width is quit different from that of the
SM. This can be clearly seen from figure 2. The enhancement of the
differential decay width increases rapidly, and the energy
spectrum of the $K^{*}$ meson is almost symmetrical. In figure 3,
the ratio $R=BR^{SM4}(B \rightarrow K^{*} \nu\bar{\nu})/BR^{SM} (B
\rightarrow K^{*} \nu\bar{\nu})$ is depicted as a function of
$(V^{*}_{\acute{t}s}V_{\acute{t}b})^{\pm}$ for various values of
$m_{\acute{t}}$. Figure 3 shows that for all values of
$m_{\acute{t}}\geq 210$ GeV the value of R becomes greater than
one. In the calculations we observed that at $m_{\acute{t}}\geq
210$, the values of R become larger than one for both solutions
$(V^{*}_{\acute{t}s}V_{\acute{t}b})^{\pm}$, meaning that the value
of R=1 is shifted. In other words, by defining the position for
which R=1, information can be obtained about $m_{\acute{t}}$ the
mass of the fourth generation fermion. For completeness we also
consider the ratio $R1=BR^{SM4}(B \rightarrow K^{*}
\nu\bar{\nu})/BR^{SM} (B \rightarrow X_{s} \nu\bar{\nu})$. This
ratio is plotted as a function of
 $(V^{*}_{\acute{t}s}V_{\acute{t}b})^{\pm}$ for various values of
$m_{\acute{t}}$ in figure 4. It is well known that the inclusive
decay width in the SM corresponds to $B \rightarrow X_{s}
\nu\bar{\nu}$ is given as (see [7]):
\begin{eqnarray}
BR(B \rightarrow X_{s}
\nu\bar{\nu})&=&\frac{3\alpha^{2}}{(2\pi)^{2}sin^{4}
\theta_{w}}\mid \frac {V_{tb}V^{*}_{ts}}{V_{cb}}\mid^{2}
\frac{[{C^{SM}}_{11}]^{2}}{\eta_{0}f(m_{c}/m_{b})}\bar{\eta}BR (B
\rightarrow X_{c} l\nu),
\end{eqnarray}
where the theoretical uncertainties related to the b-quark mass
dependence disappear. In Eq.(24) the factor 3 corresponds to the
number of the light neutrinos. Phase space factor $f(m_{c}/m_{b})
\simeq 0.44$, QCD correction factors $\eta_{0} \simeq 0.87 $,
$\bar{\eta}=1+\frac{2\alpha_{s}(m_{b})}{3\pi}
(\frac{25}{4}-\pi^{2})$$\simeq 0.83$ [9], and experimental
measurement $ BR(B \rightarrow X_{c} l\nu)=10.14\%$. Finally, note
that the results for $B\rightarrow \rho\nu\bar{\nu}$ decay can be
easily obtained from $B\rightarrow K^{*}\nu\bar{\nu}$ if the
following replacement is done in all equations: $V_{tb}V^{*}_{ts}$
by $V_{tb}V^{*}_{td}$ and $m_{K^{*}}$ by $m_{\rho}$. In obtaining
these results, one must keep in mind that the values of the form
factors for $B\rightarrow \rho$ transition generally differ from
that of the $B\rightarrow K^{*}$ transition. However, these
differences must be in the range of $SU(3)$ violation, namely in
the order $(15-20)\%$.

 \pagebreak

\pagebreak

 \begin{figure}
\centering
 \includegraphics[width=0.7\textwidth]{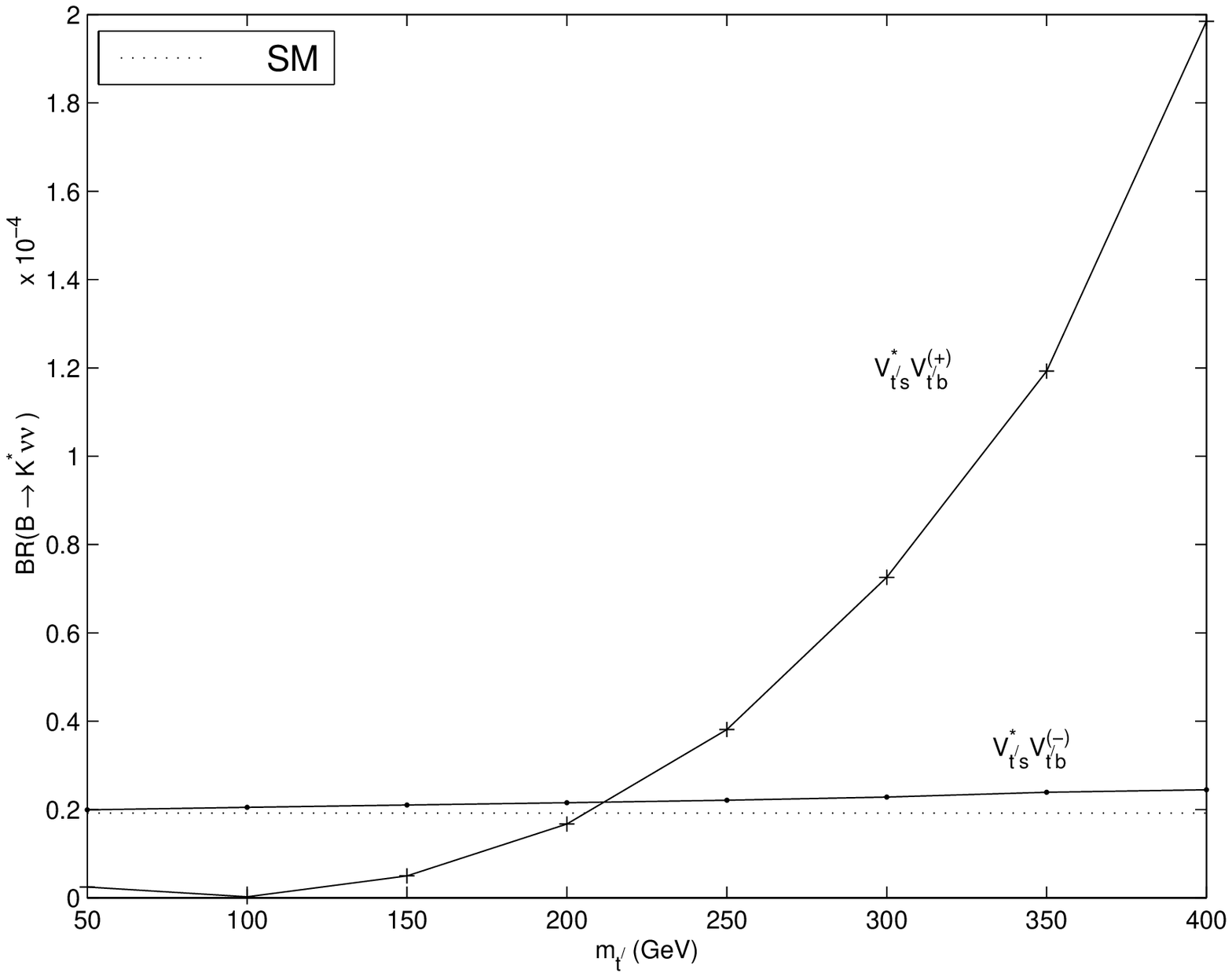}
 \caption{}
\label{fig1}
 \end{figure}
\begin{figure}
\centering
 \includegraphics[width=0.7\textwidth]{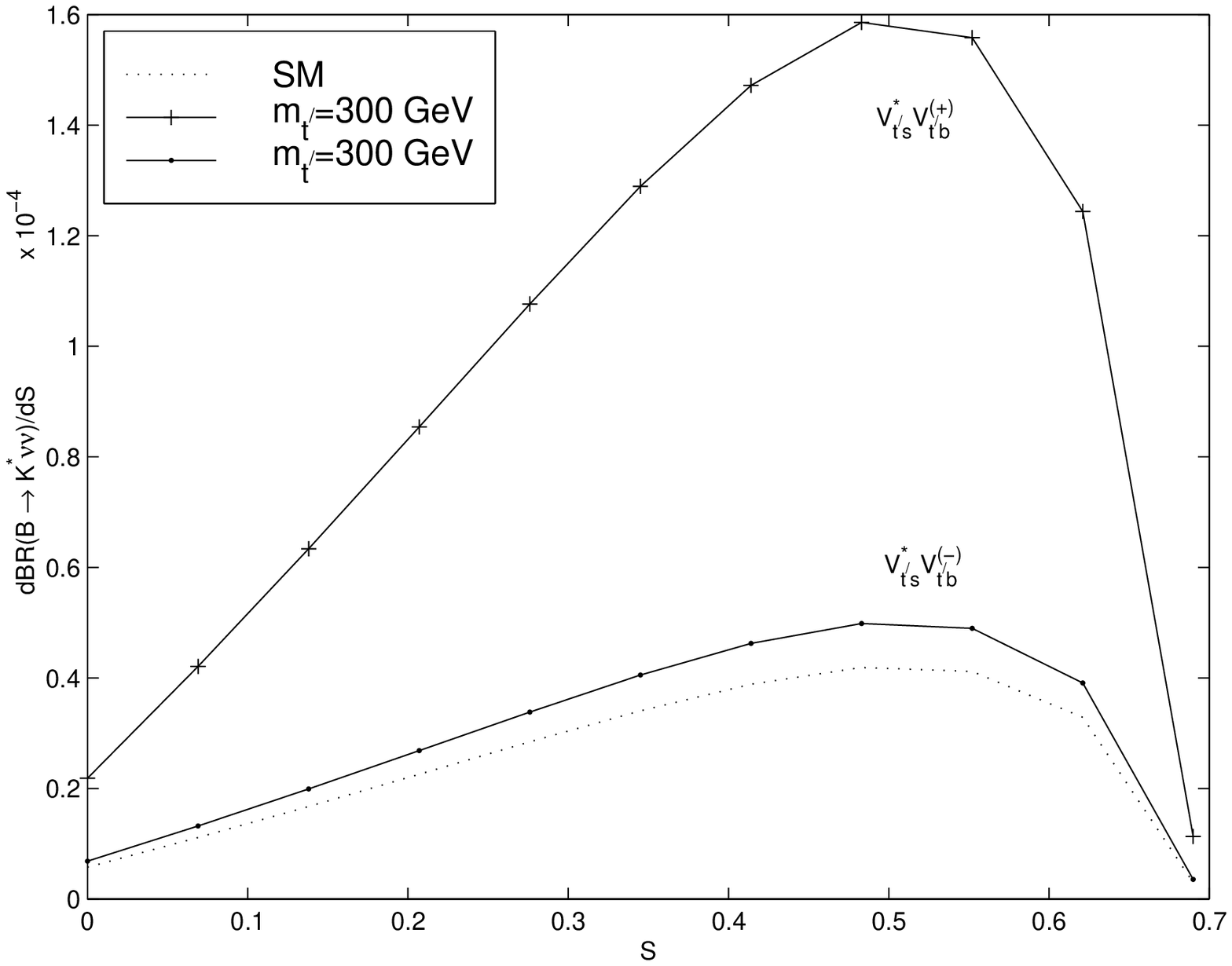}
 \caption{}
 \label{fig2}
 \end{figure}
\begin{figure}
\centering
 \includegraphics[width=0.7\textwidth]{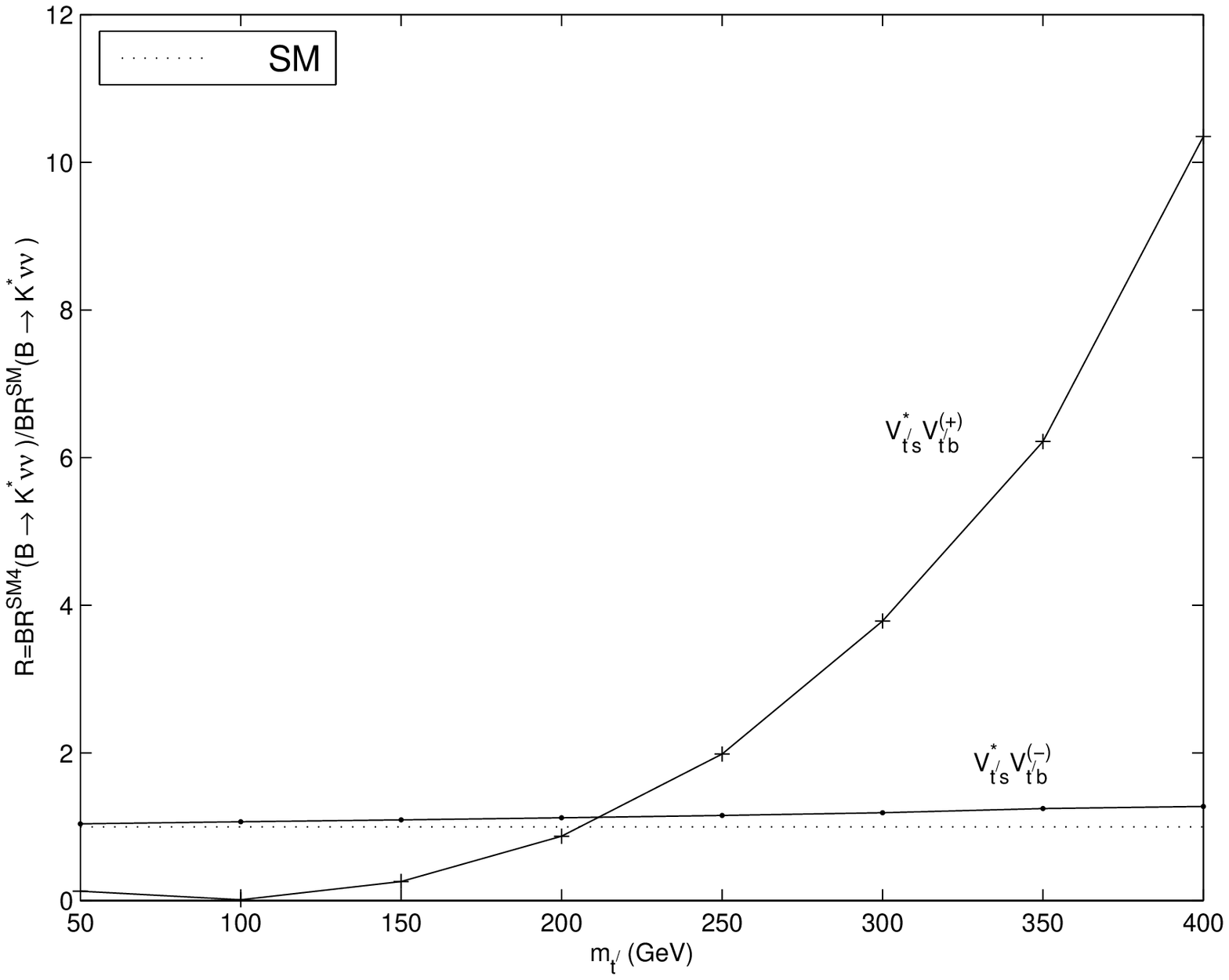}
 \caption{}
\label{fig3}
 \end{figure}
 \begin{figure}
 \centering
 \includegraphics[width=0.7\textwidth]{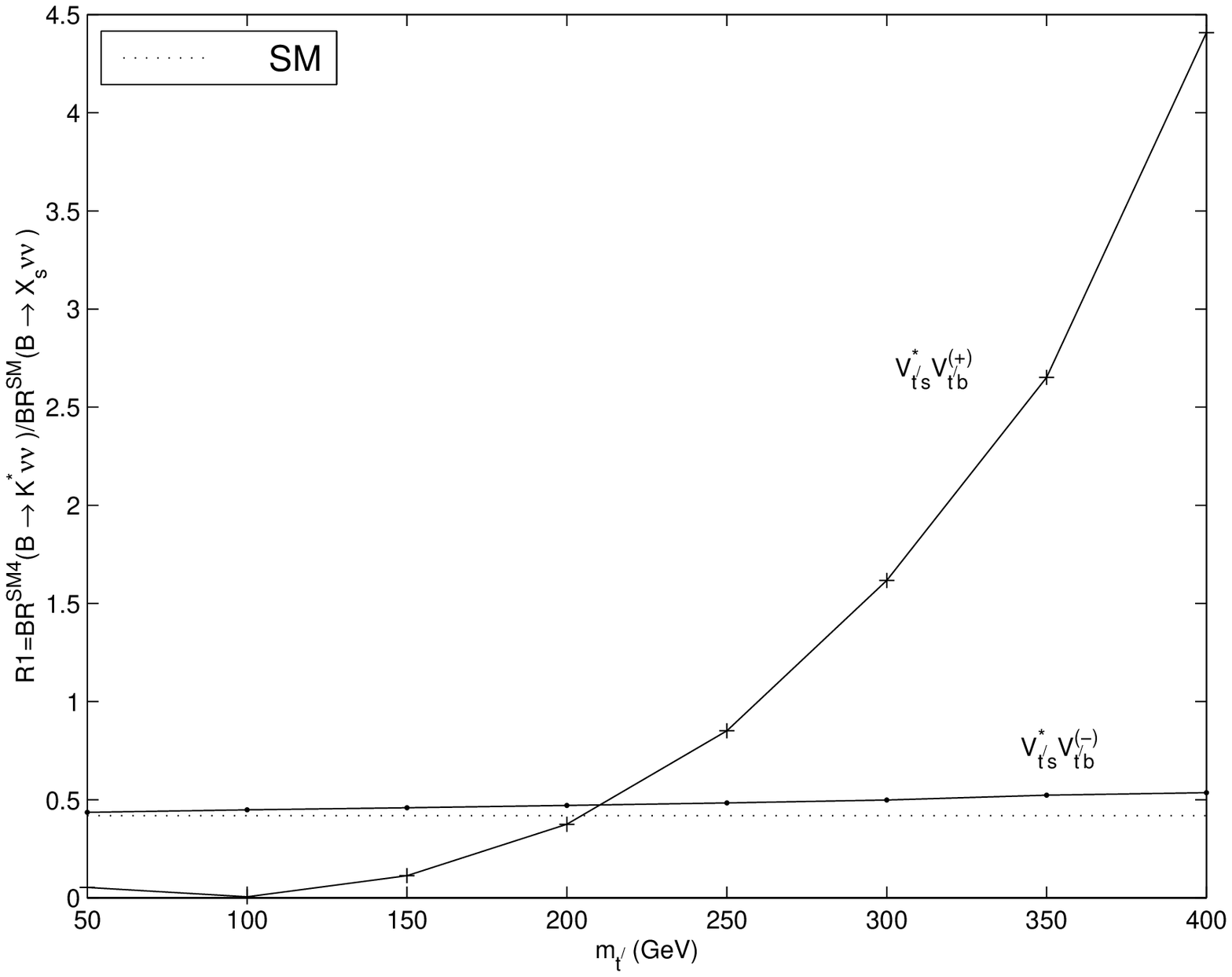}
 \caption{}
 \label{fig4}
 \end{figure}

\begin{thebibliography}{99}
\bibitem{R1}C.-S. Huang, W.-J. Huo, and Y.-L. Wu, Mod. Phys. A14
(1999) 2453, [hep-ph/9911203].
\bibitem{2} T. M. Aliev, A. $\ddot{O}$zpineci, M. Savci, Nucl. Phys. B (2000)
             275, [hep-ph/0002061].
\bibitem{3} Y. Din\c{c}er, Phys. Lett. B505, (2001) 89, [hep-ph/0012135].
\bibitem{4} K. C. Chou, Y. L. Wu, and Y. B. Xie, Chinese Phys.
Lett. 1 (1984) 2.
\bibitem{5} A. J. Buras, and M. $M\ddot{u}nz$, Phys. Rev. D52
(1995) 186, [hep-ph/9501281].
\bibitem{R6} B. Grinstein, M. J. Savage and M. B. Wise, Nucl. Phys.
B319 (1989) 271.
\bibitem{7} T. Barakat, J. Phys. G: Nucl.Part. Phys.24 (1998) 1903.
\bibitem{R8} T. Inami and C. S. Lim, Prog. Theor. Phys. 65 (1981) 287.
\bibitem{R9} G. Buchalla and A. J. Buras, Nucl. Phys. B400 (1993) 225;
             G. Buchalla and A. J. Buras and M. E. Lautenbacher,
             Rev. Mod. Phys. 68 (1996) 1125.
\bibitem{10} R. Casalbuoni et al., Phys. Reports 281 (1997) 145.
\bibitem{11} P. Colangelo, F. De Fazio, P. Santorelli, E. Scrimieri, Phys.
             Rev. D53 (1996) 3672.
\bibitem{12} W. Jaus and D. Wyler, Phys. Rev. D41 (1991) 3405; D. Melikhov,
             N. Nikitin and S. Simula, Phys. Lett. B410, (1997) 290, [hep-ph/9704268].
\bibitem{13} W. Roberts, Phys. Rev. D54 (1996) 863.
\bibitem{14} T. M. Aliev, A. $\ddot{O}$zpineci, M. Savci, Phys.Rev. D5 (1996)
             4260.
\bibitem{15} P. Ball and V. M. Braun, Phys. Rev. D55 (1997) 5561.
\bibitem{16} P. Ball, Fermilab-Conf-98/098-T, [hep-ph/9803501].
\bibitem{17} W.-J. Huo, [hep-ph/0006110].
\bibitem{18} A. J. Buras, TUM-hep-316/98, [hep-ph/9806471].
\bibitem{R19}M. S. Alam, Phys. Rev. Lett. 74 (1995) 2885.
\bibitem{R20}C. Caso et al., Particle Data Group, Eur.Phys. J. C3 (1998) 1.
\end{thebibliography}
\end{document}